\documentclass[letter,twocolumn]{jpsj3arxiv}
\usepackage{txfonts}
\usepackage[usenames,dvipsnames]{xcolor}
\usepackage{ulem}
\usepackage{bm}

\title{Topological Insulators from Electronic Superstructures}

\author{Yusuke Sugita\thanks{Email:\,sugita@aion.t.u-tokyo.ac.jp} and Yukitoshi Motome}
\inst{Department of Applied Physics, University of Tokyo, Tokyo 113-8656, Japan}

\abst{
The possibility of realizing topological insulators by the spontaneous formation of electronic superstructures is theoretically investigated in a minimal two-orbital model including both the spin-orbit coupling and electron correlations on a triangular lattice. 
Using the mean-field approximation, we show that the model exhibits several different types of charge-ordered insulators, where the charge disproportionation forms a honeycomb or kagome superstructure.
We find that the charge-ordered insulators in the presence of strong spin-orbit coupling can be topological insulators showing quantized  spin Hall conductivity. 
Their band gap is dependent on electron correlations as well as the spin-orbit coupling, and even vanishes while showing the massless Dirac dispersion at the transition to a trivial charge-ordered insulator. 
Our results suggest a new route to realize and control topological states of quantum matter by the interplay between the spin-orbit coupling and electron correlations.
}

\begin{document}
\maketitle
Quantum phenomena originating from the geometrical properties of electronic wave functions have been a central issue in modern condensed matter physics. 
The study of the anomalous Hall effect in ferromagnetic metals has revealed that the relativistic spin-orbit coupling (SOC) plays a crucial role in such phenomena~\cite{RevModPhys.82.1539}.
In particular, the strong SOC may bring about intriguing topological states of matter, such as the topological insulator (TI)~\cite{RevModPhys.82.3045,RevModPhys.83.1057,doi:10.7566/JPSJ.82.102001}. 
The TI is a nontrivial band insulator, which is distinguished from conventional ones by a $Z_{2}$ topological invariant under the time-reversal symmetry.
It exhibits a peculiar metallic edge (or surface) state, which gives rise to the quantized spin Hall effect in two-dimensional TIs.
Such an unusual edge or surface state has been observed experimentally in several systems, e.g., a two-dimensional quantum well of CdTe/HgTe/CdTe~\cite{Konig766} and three-dimensional bulk crystals of Bi$_{x}$Sb$_{1-x}$~\cite{hsieh2008topological}.

Recently, electron correlations in the systems with strong SOC have attracted much interest.
In weakly correlated systems, the band topology survives and the spontaneous symmetry breaking by electron correlations may lead to new types of topological phases, such as Weyl semimetals by spatial-inversion or time-reversal symmetry breaking~\cite{doi:10.1146/annurev-conmatphys-020911-125138,doi:10.1146/annurev-conmatphys-031113-133841,doi:10.7566/JPSJ.83.061017}.
On the other hand, strong electron correlations in the presence of strong SOC give rise to highly anisotropic exchange interactions in the Mott insulator, which may lead to unconventional quantum phases, e.g., quantum spin liquids~\cite{doi:10.1146/annurev-conmatphys-020911-125138,doi:10.1146/annurev-conmatphys-031115-011319}.
Thus, the interplay between the SOC and electron correlations provides a key for new quantum phenomena, but the survey has only been initiated and many aspects remain unexplored.

In this Letter, we propose a new route to realize topological states of matter through the interplay between the SOC and electron correlations. 
In our scenario, spontaneous symmetry breaking takes place to form an electronic superstructure, such as a charge density wave, which brings about a topological nature in the band structure. 
We examine this scenario in a minimal model on a triangular lattice, which mimics some delafossite-type  oxides~\cite{ong2004electronic,PhysRevLett.99.157204} and transition metal dichalcogenides~\cite{0953-8984-23-21-213001,chhowalla2013chemistry}.
Using the mean-field approximation,  we clarify the ground-state phase diagram at commensurate electron fillings while changing the SOC and electron correlations.
We find that the system becomes TIs, in some specific charge-ordered states, where the charge disproportionation comprises a honeycomb or kagome superstructure.
We show that such charge-ordered TIs are stabilized by the cooperation of the SOC and electronic correlations.    
Our results indicate the new possibility of realizing and controlling TIs through electronic superstructures. 

\begin{figure}
\centering
\includegraphics[width=0.48\textwidth]{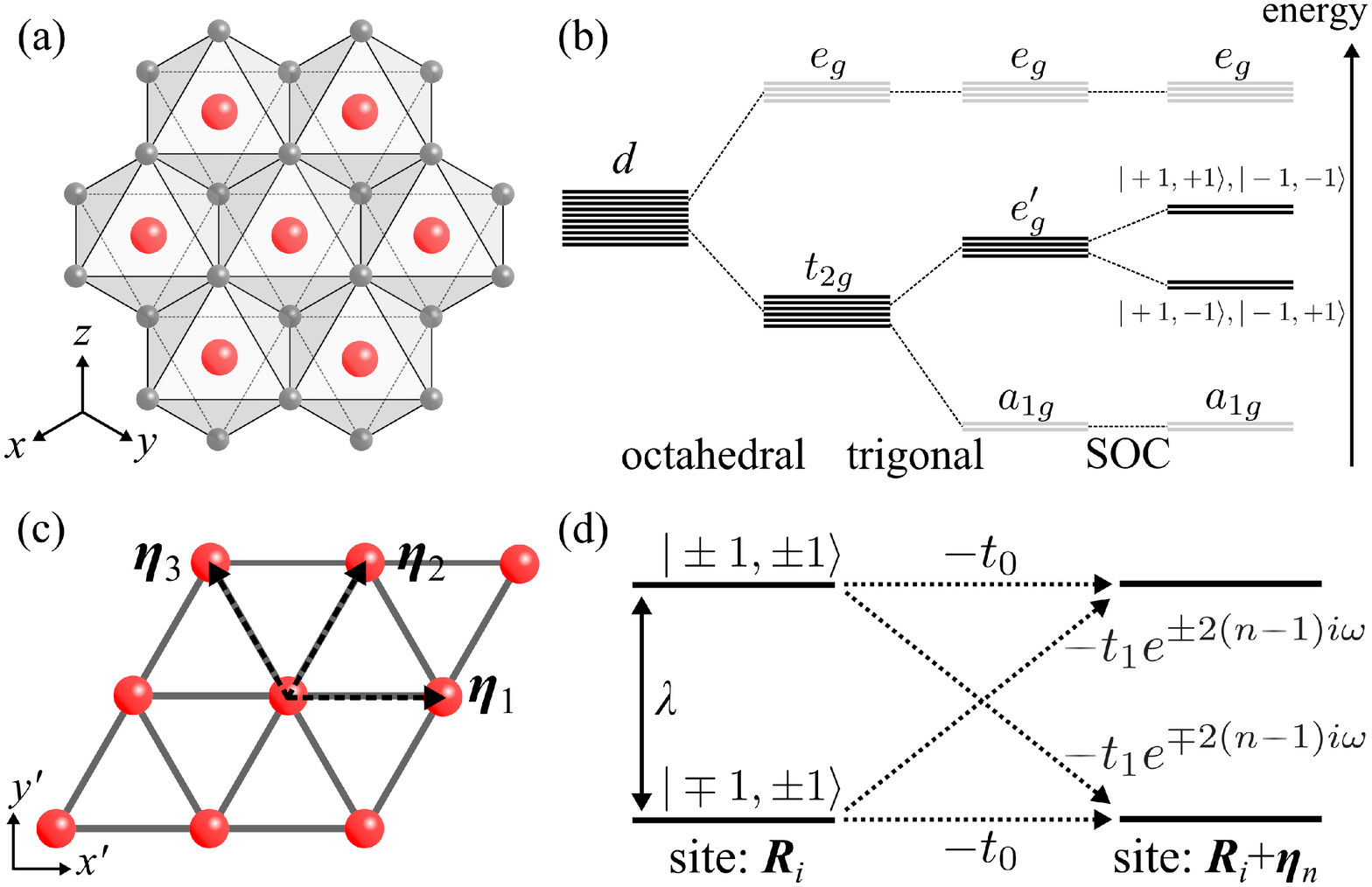}
\caption{
(Color online)
(a) Schematic picture of edge-sharing octahedra.
The large (red) spheres inside the octahedra denote the transition metal cations and the small (gray) ones on the vertices indicate the ligand ions.
(b) Atomic $d$ orbital levels of  the transition metal cations under the octahedral and trigonal crystalline electric fields corresponding to (a).
The $e'_{g}$ orbitals are further split by the SOC; see the text for details.  
(c) Schematic picture of the triangular lattice of transition metal cations. 
$\bm{\eta}_n$ ($n=1,2,3$) are the primitive translational vectors. 
(d) Energy levels and hopping processes in the two-orbital model in Eq. (\ref{ModOne}). 
}
\label{Setup}
\end{figure}

To investigate the spontaneous formation of electronic superstructures and resultant topological nature, we consider a minimal model on a triangular lattice. 
We begin with the edge-sharing octahedra composed of transition metal cations and ligands, as shown in Fig.~\ref{Setup}(a). 
Note that a similar situation is realized in delafossite compounds~\cite{ong2004electronic,PhysRevLett.99.157204} and 1T-type transition metal dichalcogenides~\cite{0953-8984-23-21-213001,chhowalla2013chemistry}.
When the octahedral and trigonal crystalline electric fields are sufficiently large, the $d$ orbitals in the transition metal cations are split into three groups, $e_g$, $e'_g$, and $a_{1g}$, as shown in Fig.~\ref{Setup}(b).
Assuming that the Fermi level is at the $e'_g$ manifold (otherwise, the SOC is rather irrelevant), we take into account only the $e'_g$ orbitals and omit the others. 
The $e'_g$ states are denoted by $|m=\pm1,\sigma\rangle = (|xy,\sigma\rangle + e^{\pm i\omega}|yz,\sigma\rangle + e^{\mp i\omega}|zx,\sigma\rangle)/\sqrt{3}$, where $\omega = 2\pi/3$ and $xyz$-axes are taken as shown in Fig.~\ref{Setup}(a); $\sigma=\pm1$ denotes the spin, whose quantization axis is taken along the (111) direction.
Under these assumptions, we construct a tight-binding model for the triangular lattice composed of transition metal cations with $e'_g$ orbitals, whose one-body Hamiltonian is given by 
\begin{align}
H_{0}
=
&-t_{0}\sum_{\bm{k}}\sum_{m,\sigma=\pm1}
\gamma_{0\bm{k}}c^{\dagger}_{\bm{k}m\sigma}c_{\bm{k}m\sigma}
-t_{1}\sum_{\bm{k}}\sum_{m,\sigma=\pm1}
\gamma_{m\bm{k}}c^{\dagger}_{\bm{k}m\sigma}c_{\bm{k}-m\sigma} \nonumber \\
&+\frac{\lambda}{2}\sum_{\bm{k}}\sum_{m,\sigma=\pm1}
(m\sigma)c^{\dagger}_{\bm{k}m\sigma}c_{\bm{k}m\sigma},
\label{ModOne}
\end{align}
where $c^{\dagger}_{\bm{k} m \sigma}$($c_{\bm{k} m \sigma}$) is the creation (annihilation) operator of an electron for the wave vector $\bm{k}$, orbital $m = \pm1$, and spin $\sigma =\pm1$. 
$t_{0}$ and $t_{1}$ are the intra- and interorbital hopping elements between nearest-neighbor sites, respectively.
The factors $\gamma_{\alpha\bm{k}}$ ($\alpha=0,\pm1$) in the hopping terms in Eq.~(\ref{ModOne}) are given by
\begin{align}
\gamma_{\alpha\bm{k}} = \sum_{n=1,2,3} 2 e^{2(n-1)i \alpha\omega}\cos{(\bm{k}\cdot \bm{\eta}_{n})},
\label{ModGam}
\end{align}
which originate from the directional dependences of the overlaps between $|xy \rangle$, $|yz \rangle$, and $|zx \rangle$ orbitals.
Hereafter, we take the triangular plane as the $x'y'$ plane, and set the primitive translational vectors for the triangular lattice as $\bm{\eta}_{1} = (1, 0)$, $\bm{\eta}_{2} = (1/2, \sqrt{3}/2)$, and $\bm{\eta}_{3} = (-1/2, \sqrt{3}/2)$ [see Fig.~\ref{Setup}(c)].
$\lambda$ is the SOC constant, which splits the energy levels of $|m,\sigma\rangle$ into two Kramers doublets, $\{|+1,+1\rangle,|-1,-1\rangle\}$ and $\{|+1,-1\rangle,|-1,+1\rangle\}$, when $t_0=t_1=0$, as shown in Fig.~\ref{Setup}(b).
The energy levels and hopping processes are schematically shown in Fig.~\ref{Setup}(d).
Note that a similar model was studied on a honeycomb lattice~\cite{PhysRevB.90.081115,1742-6596-592-1-012131}.
Although the honeycomb lattice model exhibits topologically nontrivial states even in the noninteracting case~\cite{1742-6596-592-1-012131}, our triangular lattice model in Eq.~(\ref{ModOne}) does not for any values of the parameters $t_{0}$, $t_{1}$, and $\lambda$.
Hereafter, we set $t_{0}=0.5$ and $t_{1}=0.25$, which are reasonable when considering the $d$-$d$ direct and $d$-$p$-$d$ indirect hoppings in the Slater--Koster scheme~\cite{PhysRev.94.1498}.

In addition to the one-body part, we take into account both the onsite and intersite Coulomb interactions.
The onsite one is given by
\begin{equation}
H^{\rm onsite}_{1} =
\frac{1}{2} \sum_{mnm'n'} U_{mnm'n'}
\sum_{i}
\sum_{\sigma \sigma'}
c^{\dagger}_{i m\sigma} c^{\dagger}_{i n\sigma'} c_{i n'\sigma'} c_{im'\sigma}.
\label{ModOns}
\end{equation}
Assuming the rotational symmetry of the Coulomb interaction, we set $U_{mmmm} = U$,  $U_{mnmn} = U-2J$, and $U_{mnnm}=U_{mmnn}=J$ ($m\neq n$), where $U$ is the intraorbital Coulomb interaction and $J$ is the Hund's coupling, respectively.
We set $U=1.0$ and $J/U=0.1$ in the following calculations. 
For the intersite interaction, we consider the density-density repulsions given by
\begin{equation}
H^{\rm intersite}_{1}
= 
V_1 \sum_{\langle i,j \rangle} n_{i}n_{j} + V_2\sum_{\langle\langle i,j \rangle\rangle} n_{i}n_{j},
\label{ModInt}
\end{equation}
where $n_{i} = \sum_{m\sigma}c^{\dagger}_{i m\sigma} c_{i m\sigma}$. 
The sum of $\langle i,j \rangle$ ($\langle\langle i,j \rangle\rangle$) is taken for the nearest- (next-nearest-) neighbor sites.

\begin{figure}
\centering
\includegraphics[width=0.48\textwidth]{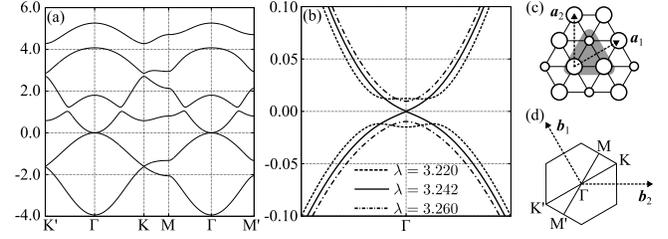}
\caption{
(a) Electronic band structure for the type-B COI at $V_{1}=0.6$ and $\lambda=3.242$.
(b) Enlarged figure of the band structures near the Fermi level around the $\Gamma$ point along the K'-K line, for $V_{1}=0.6$ and $\lambda=3.220$, $3.242$, and $3.260$. 
(c) Schematic picture of the three-site unit cell (gray triangle) used for drawing the electronic band structures.
$\bm{a}_{n}$ ($n=1,2$) are the primitive translational vectors.
(d) Schematic picture of the folded Brillouin zone for the unit cell in (c).
$\bm{b}_{n}$ ($n=1,2$) are the reciprocal lattice vectors.
}
\label{BandHon}
\end{figure}

\begin{figure}
\centering
\includegraphics[width=0.48\textwidth]{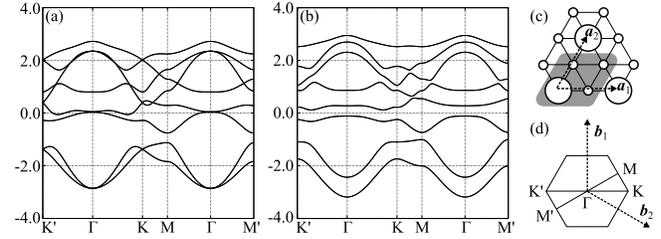}
\caption{
(a) and (b) Electronic band structure for the kagome CO phase at $V_{1}=0.6$, $V_{2}=0.15$, and (a) $\lambda=0.0$ and (b) $\lambda=0.6$.
(c) Schematic picture of the unit cell used for drawing the electronic band structures of the kagome CO phase.
The gray region indicates the unit cell composed of four sites.
$\bm{a}_{n}$ ($n=1,2$) are the primitive translational vectors.
(d) Schematic picture of the folded Brillouin zone for the unit cell in (c).
$\bm{b}_{n}$ ($n=1,2$) are the reciprocal lattice vectors.
}
\label{BandKag}
\end{figure}

To clarify the ground states of the two-orbital model given by Eqs.~(\ref{ModOne}), (\ref{ModOns}), and (\ref{ModInt}), we use the mean-field approximation.
In the mean-field calculation, we employ 12 sublattices [see Figs.~\ref{BandHon}(c) and \ref{BandKag}(c)] and approximate the integration in the folded Brillouin zone by the summation over $64\times64$ $\bm{k}$ points. 
We apply the Hartree-Fock approximation to the onsite interaction in Eq.~(\ref{ModOns}) and the Hartree approximation to the intersite interaction in Eq.~(\ref{ModInt}) to focus on charge order\cite{note1}.
The mean fields are determined self-consistently, until they converge within a precision of less than $10^{-6}$.
We investigate the ground state and find several interesting charge orders at nearly 1/3 electron filling, $\langle \sum_i n_i \rangle/(4N) \sim 1/3$, where $N$ is the number of lattice sites.

In addition, we compute the spin Hall conductivity, which signals the nontrivial topological nature of the system; it can be quantized at a nonzero integer multiple value of $e/2\pi$ for two-dimensional TIs~\cite{RevModPhys.82.3045} ($e$ is the elementary charge).
Using the standard Kubo formula in the linear response theory, we calculate the spin Hall conductivity as
\begin{equation} 
\sigma^{s}_{xy}
=
\frac{e}{2}\frac{1}{i\Omega}
\sum_{\bm{k} \alpha\beta}
\frac{f(\varepsilon_{\alpha\bm{k}}) - f(\varepsilon_{\beta\bm{k}})}{\varepsilon_{\alpha\bm{k}} - \varepsilon_{\beta\bm{k}}}
\frac{\langle \alpha\bm{k}| j^{s}_{x} |\beta\bm{k} \rangle \langle \beta\bm{k}| j_{y} |\alpha\bm{k} \rangle}{\varepsilon_{\alpha\bm{k}}-\varepsilon_{\beta\bm{k}}+i\delta},
\label{SHC}
\end{equation}
where $\Omega$ is the system volume, $f$ is the Fermi distribution function, $\varepsilon_{\alpha\bm{k}}$ and $|\alpha\bm{k} \rangle$ are the eigenvalues and eigenvectors of the $\alpha$th electronic band with the wave vector $\bm{k}$ in the mean-field solution, respectively, and $\delta$ is the infinitesimal positive parameter. 
In the calculation of Eq.~(\ref{SHC}), we take the summation over $512\times512$ $\bm{k}$ points and set $T=10^{-3}$ and $\delta=10^{-3}$. 
The current operator is defined by $j_{y} \equiv \partial H_{\rm MF}/\partial k_{y}$, where $H_{\rm MF}$ is the mean-field Hamiltonian.
We define the spin current operator as $j^{s}_{x} \equiv \{ j_{x}, \sigma _{z} \}/2$, where $\sigma_z$ is the $z$-component of the Pauli matrices for spin; 
note that $H_{\rm MF}$ for our mean-field solutions commutes with $\sigma_z$. 
Hereafter, we denote $\tilde{\sigma}^{s}_{xy} \equiv \sigma^{s}_{xy}/(e/2\pi)$ as the normalized spin Hall conductivity.
We note that, for two-dimensional systems whose Hamiltonian commutes with $\sigma_z$, $\tilde{\sigma}^{s}_{xy}$ is directly related to the $Z_{2}$ topological invariant~\cite{RevModPhys.82.3045}.

\begin{figure}
\centering
\includegraphics[width=0.48\textwidth]{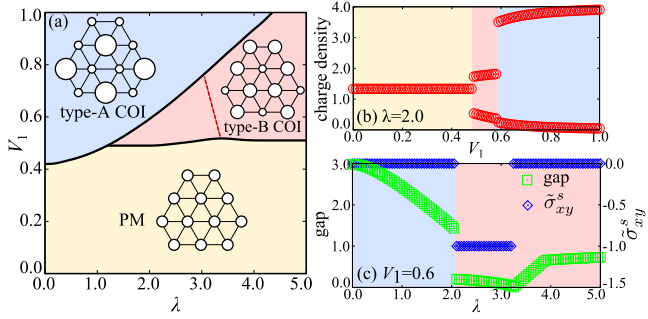}
\caption{
(Color online)
(a) Ground-state phase diagram for the model given by Eqs.~(\ref{ModOne}), (\ref{ModOns}), and (\ref{ModInt}) at 1/3 filling obtained by the mean-field approximation.
We set $U=1.0$, $J/U=0.1$, and $V_2=0$. 
A schematic picture of the charge ordering pattern is shown in each phase.
The size of the circle represents the magnitude of the local charge density at each sublattice.
In the type-B COI phase, the band gap closes on the red dashed line, which corresponds to the phase boundary between the TI and trivial band insulator.   
(b) $V_1$ dependence of the local charge density at each sublattice at $\lambda=2.0$.
(c) $\lambda$ dependences of the band gap and the normalized spin Hall conductivity at $V_{1}=0.6$.
}
\label{MF1/3}
\end{figure}

First, we consider the situation where $V_{1}$ is dominant rather than $V_{2}$, and thus, set $V_{2}=0$. 
In this case, the system shows an interesting behavior at 1/3 filling. 
Figure~\ref{MF1/3}(a) shows the ground-state phase diagram obtained by the mean-field approximation while changing $\lambda$ and $V_{1}$.
We find three different phases in this parameter region: paramagnetic metal (PM) for small $V_1$ and two charge-ordered insulators (COIs) for large $V_1$.
In both COIs, the local charge density is disproportionated to form a honeycomb superstructure; in the COI in the larger $V_1$ region, the local charge density is lower at the sites belonging to the honeycomb network than at the isolated sites, while they are opposite in the COI in the intermediate $V_1$ and large $\lambda$ region [see the schematic pictures in Fig.~\ref{MF1/3}(a)]. 
The local charge densities are plotted in Fig.~\ref{MF1/3}(b). 
We call the former (latter) the honeycomb type-A(B) COI. 

In the small $\lambda$ region, there is a transition from the paramagnetic metal to the honeycomb type-A COI with increasing $V_1$. 
This is easily understood by considering that the electrons tend to avoid each other under large $V_1$ and the lowest energy configuration at 1/3 filling is given by the type-A charge ordering.
On the other hand, in the large $\lambda$ region, the type-B COI appears between the type-A COI and PM phases.
The intervening type-B COI is stabilized by the synergy between the strong SOC and intersite Coulomb repulsion. 
This is understood by considering the large $\lambda$ limit as follows. 
As the bands for the Kramers pair are largely split from each other, the two-orbital model at 1/3 filling reduces to a single-band model at 2/3 filling for the lower-energy band.
In the single-band model at 2/3 filling, the lowest energy state under $V_1$ is given by the type-B charge ordering, which explains why the type-B COI is stabilized in the large $\lambda$ and $V_1$ region in Fig.~\ref{MF1/3}(a).
We note that all the phase boundaries in Fig.~\ref{MF1/3}(a) are of first order with discontinuous changes in local charge densities.

Figure~\ref{MF1/3}(c) shows $\lambda$ dependences of the energy gap and normalized spin Hall conductivity at $V_1=0.6$.
The results clearly indicate that both CO states are gapped insulators, while the type-A COI has a larger gap than the type-B COI in this parameter region. 
Remarkably, in the type-B CO state, the band gap once closes around $\lambda \sim 3.2$.
The gapless line inside the type-B CO phase is shown by the dashed line in Fig.~\ref{MF1/3}(a). 
The result indicates that the type-B CO phase may include two different insulating states separated by the gapless boundary.
Indeed, as shown in Fig.~\ref{MF1/3}(c), the normalized spin Hall conductivity in the type-B COI is quantized at $-1$ for $\lambda \lesssim 3.2$, while it changes discontinuously to zero when crossing the gapless point.
Therefore, the gapless boundary in the type-B COI corresponds to a topological transition between a TI for smaller $\lambda$ and a trivial band insulator for larger $\lambda$.

To clarify the electronic states in the type-B COI further, we show the electronic band structure of the mean-field solution for the type-B COI near the gapless boundary in Figs.~\ref{BandHon}(a) and \ref{BandHon}(b) [the unit cell and Brillouin zone are shown in Figs.~\ref{BandHon}(c) and \ref{BandHon}(d), respectively].
As shown in Fig.~\ref{BandHon}(a), the Kramers doublets are split by the strong SOC into two `copies' of three bands; the highest band in the lower three bands hybridizes with the lowest one in the higher three bands, resulting in a small gap at $\varepsilon\sim1.1$. 
The three bands in each copy are composed of two subsets, reflecting the honeycomb CO superstructure; the lower two bands comprise the dispersive bands similar to those of the single-band model on the honeycomb lattice, and the remaining higher band is less dispersive as it comes from the isolated sites in the honeycomb hexagons.
The lower honeycomb-like bands are occupied (the Fermi level is set at zero). 
This result supports the above discussion for the origin of the type-B COI.

Figure \ref{BandHon}(b) shows more details of the band structures near the Fermi level at 1/3 filling around the $\Gamma$ point at $V_{1}=0.6$.
With increasing $\lambda$, the band gap at 1/3 filling decreases and closes at $\lambda\sim3.242$.
In the gapless state, the low-energy dispersions are well approximated by the massless Dirac cone.
The Dirac cone is gapped out again by further increasing $\lambda$.
We note that, although this topological transition appears to share the fundamental mechanism with that found for the similar two-orbital model on a honeycomb lattice~\cite{1742-6596-592-1-012131}, the critical value of $\lambda$ is largely reduced by the mean-field contribution from electron correlations.
In other words, electron correlations enhance the effective SOC for realizing the topological state of matter.

\begin{figure}
\centering
\includegraphics[width=0.48\textwidth]{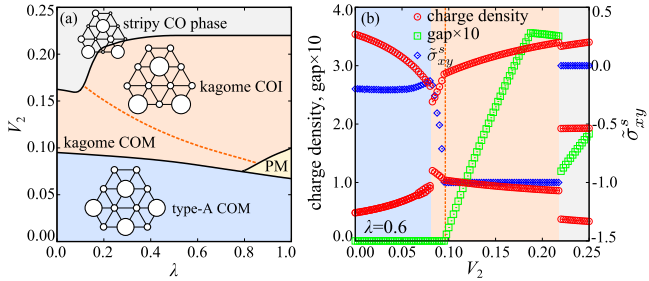}
\caption{
(Color online)
(a) Ground-state phase diagram for the model given by Eqs.~(\ref{ModOne}), (\ref{ModOns}), and (\ref{ModInt}) at 3/8 filling obtained by the mean-field approximation.
We set $U=1.0$, $J/U=0.1$, and $V_{1}=0.6$. 
Schematic picture of the charge ordering pattern is shown in each phase.
The size of the circle represents the magnitude of the local charge density at each sublattice.  
In the kagome CO phase, the orange dashed line separates the metallic and insulating regions.   
(b) $V_2$ dependences of the charge density at each sublattice, the band gap, and the normalized spin Hall conductivity at $\lambda=0.6$.
}
\label{MF3/8}
\end{figure}

Next, we take into account the next-nearest-neighbor repulsion $V_{2}$.
We find that $V_2$ leads to different types of electronic superstructures around 1/3 filling. 
In particular, here, we discuss interesting CO states appearing at 3/8 filling.
Figure~\ref{MF3/8}(a) shows the ground-state phase diagram at $V_{1}=0.6$ while changing $\lambda$ and $V_{2}$.
In the small $V_{2}$ region, the system exhibits a honeycomb type-A CO metal (COM) as well as PM, whose charge patterns are also seen in the 1/3 filling case above~\cite{note2}.
When increasing $V_{2}$, we find two new CO phases: kagome and stripy CO phases [see the schematic picture in the phase diagram in Fig.~\ref{MF3/8}(a)].
In the kagome CO state, the charge density is disproportionated so that the charge-poor sites comprise a kagome superstructure, as plotted in Fig.~\ref{MF3/8}(b). 
(The charge-poor sites have a very small charge disproportionation among them, which does not affect the following topological nature of this phase.)
On the other hand, the stripy CO state has a four-sublattice order, where the charge density is disproportionated into three groups: charge-rich, charge-poor, and intermediate at one, two, and one sublattices, respectively [see Fig.~\ref{MF3/8}(b)].

The kagome CO state is intriguing from the topological viewpoint, as discussed below. 
In Fig.~\ref{MF3/8}(b), we plot the $V_2$ dependence of the energy gap at $\lambda=0.6$.
The result shows that the kagome CO phase is metallic in the small $V_2$ region but becomes insulating with increasing $V_2$.
The band gap is opened by the cooperation between the SOC and $V_{2}$ [see also Fig.~\ref{MF3/8}(a)].
This is explicitly shown in the band structures for $V_2=0.15$ in Figs.~\ref{BandKag}(a) and \ref{BandKag}(b) at $\lambda=0.0$ and $\lambda=0.6$, respectively [the unit cell and Brillouin zone are shown in Figs.~\ref{BandKag}(c) and \ref{BandKag}(d), respectively].
Although the bands near the Fermi level do not have a gap at $\lambda=0.0$, they are separated by a gap for $\lambda=0.6$.
We find that the kagome COI is a TI by calculating the normalized spin Hall conductivity, as shown in Fig.~\ref{MF3/8}(b).
Although the normalized spin Hall conductivity is already nonzero in the honeycomb type-A and kagome COM phases for smaller $V_2$, it is quantized at a nonzero integer number, $\tilde{\sigma}^{s}_{xy}=-1.0$ in the kagome COI. 

Finally, let us discuss our results.
We found two different types of COIs which are topologically nontrivial: the honeycomb type-B and kagome COIs. 
The important physics here is the role of the SOC under the electronic superstructures.
As remarked above, the noninteracting model including the SOC [Eq.~(\ref{ModOne})] does not exhibit any topological nature owing to the high symmetry of the triangular lattice.
The formation of the honeycomb and kagome superstructures activates the hidden SOC effect and changes the system into TIs.
This is, for instance, understood from the relationship between the electronic states of the honeycomb type-B COI in the present model and the PM in the honeycomb-lattice model studied in a previous work~\cite{1742-6596-592-1-012131}, as discussed above.
The situation is distinct from other interaction-driven TIs, the so-called topological Mott insulators~\cite{doi:10.7566/JPSJ.83.061017,PhysRevLett.100.156401,PhysRevB.82.045102,PhysRevB.82.075125}, where the atomic SOC does not play an important role~\cite{note1}.

Similar mechanisms activating the SOC effect by superstructure formation were discussed for spatial inversion symmetry breaking, which induces the antisymmetric SOC~\cite{PhysRevB.90.081115,doi:10.7566/JPSJ.83.014703,doi:10.7566/JPSJ.83.114704}.
Thus, our results point to a much broader route to activate the nontrivial SOC physics and realize topological states of matter, with the aid of the change of spatial symmetry by electronic correlations.
This has richer implications, since, in addition to charge ordering, the superstructure formation can be caused by other degrees of freedom, e.g., magnetic ordering in spin-charge coupled systems~\cite{PhysRevLett.109.237207,PhysRevB.88.100402,PhysRevB.91.155132} and bond ordering in electron-phonon coupled systems.
Interestingly, there are many candidate materials exhibiting various superstructures, e.g., delafossite-type oxides~\cite{ong2004electronic,PhysRevLett.99.157204} and transition metal dichalcogenides~\cite{0953-8984-23-21-213001,chhowalla2013chemistry}. 
In particular, the latter compounds are intriguing, as they show a variety of charge density waves with longer periodicities accompanied by lattice distortions. 
The topological nature in these interesting states with electronic superstructures is left for a future study.

\begin{acknowledgment}
The authors thank Satoru Hayami, Hiroaki Kusunose, Takahiro Misawa, and Youhei Yamaji for constructive suggestions. 
Y.S. is supported by the Japan Society for the Promotion of Science through the Program for Leading Graduate Schools (MERIT).
This work was supported by Grants-in-Aid for Scientific Research (Nos. 24340076 and 15K05176), the Strategic Programs for Innovative Research (SPIRE), MEXT, and the Computational Materials Science Initiative (CMSI), Japan.
\end{acknowledgment}


\begin{thebibliography}{27}
\bibitem{RevModPhys.82.1539}
N.~Nagaosa, J.~Sinova, S.~Onoda, A.~H.~MacDonald, and N.~P.~Ong, Rev.~Mod.~Phys. {\bf 82}, 1539 (2010).
\bibitem{RevModPhys.82.3045}
M.~Z.~Hasan and C.~L.~Kane, Rev.~Mod.~Phys. {\bf 82}, 3045 (2010).

\bibitem{RevModPhys.83.1057}
X.-L.~Qi and S.-C.~Zhang, Rev.~Mod.~Phys. {\bf 83}, 1057 (2011).

\bibitem{doi:10.7566/JPSJ.82.102001}
Y.~Ando, J.~Phys.~Soc.~Jpn. {\bf 82}, 102001 (2013).
\bibitem{Konig766}
M.~K{\"o}nig, S.~Wiedmann, C.~Br{\"u}ne, A.~Roth, H.~Buhmann, L.~W.~Molenkamp, X.-L.~Qi, and S.-C.~Zhang, Science {\bf 318}, 766 (2007).
\bibitem{hsieh2008topological}
D.~Hsieh, D.~Qian, L.~Wray, Y.~Xia, Y.~S.~Hor, R.~J.~Cava, and M.~Z.~Hasan, Nature {\bf 452}, 970 (2008).
\bibitem{doi:10.1146/annurev-conmatphys-020911-125138}
W.~Witczak-Krempa, G.~Chen, Y.~B.~Kim, and L.~Balents, Annu.~Rev.~Condens.~Matter~Phys. {\bf 5}, 57 (2014).

\bibitem{doi:10.1146/annurev-conmatphys-031113-133841}
O.~Vafek and A.~Vishwanath, Annu.~Rev.~Condens.~Matter~Phys. {\bf 5}, 83 (2014).

\bibitem{doi:10.7566/JPSJ.83.061017}
M.~Imada, Y.~Yamaji, and M.~Kurita, J.~Phys.~Soc.~Jpn. {\bf 83}, 061017 (2014).

\bibitem{doi:10.1146/annurev-conmatphys-031115-011319}
J.~G.~Rau, E.~K.-H.~Lee, and H.-Y.~Kee, Annu.~Rev.~Condens.~Matter~Phys. {\bf 7}, 195 (2016).
\bibitem{ong2004electronic}
N.~P.~Ong and R.~J.~Cava, Science {\bf 305}, 52 (2004).

\bibitem{PhysRevLett.99.157204}
E.~Wawrzy\'{n}ska, R.~Coldea, E.~M.~Wheeler, I.~I.~Mazin, M.~D.~Johannes, T.~S\"orgel, M.~Jansen, R.~M.~Ibberson, and P.~G.~Radaelli, Phys.~Rev.~Lett. {\bf 99}, 157204 (2007).
\bibitem{0953-8984-23-21-213001}
K.~Rossnagel, J.~Phys.:~Condens.~Matter {\bf 23}, 213001 (2011).

\bibitem{chhowalla2013chemistry}
M.~Chhowalla, H.~S.~Shin, G.~Eda, L.-J.~Li, K.~P.~Loh, and H.~Zhang, Nat.~Chem. {\bf 5}, 263 (2013).
\bibitem{PhysRevB.90.081115}
S.~Hayami, H.~Kusunose, and Y.~Motome, Phys.~Rev.~B {\bf 90}, 081115 (2014).

\bibitem{1742-6596-592-1-012131}
S.~Hayami, H.~Kusunose, and Y.~Motome, J.~Phys.:~Conf.~Ser. {\bf 592}, 012131 (2015).
\bibitem{PhysRev.94.1498}
J.~C.~Slater and G.~F.~Koster, Phys.~Rev. {\bf 94}, 1498 (1954).
\bibitem{note1}
Imaginary terms from the Fock decoupling of the intersite interactions may prefer an interaction-driven TI, the so-called topological Mott insulator\cite{doi:10.7566/JPSJ.83.061017,PhysRevLett.100.156401,PhysRevB.82.075125,PhysRevB.82.045102}.
In this study, however, we focus on the charge ordering as a trigger for TIs.
We confirmed that the result remains qualitatively the same, when adopting the three-site unit cell as shown in Fig.~\ref{BandHon}(c) and the Hartree-Fock approximation for the intersite interactions in the case of 1/3 filling.
\bibitem{note2}
In the type-A COM, the charge density is slightly modulated in a staggered manner on the honeycomb network near the phase boundary to the kagome CO state [see also Fig.~\ref{MF3/8}(b)].
\bibitem{PhysRevLett.100.156401}
S.~Raghu, X.-L.~Qi, C.~Honerkamp, and S.-C.~Zhang, Phys.~Rev.~Lett. {\bf 100}, 156401 (2008).

\bibitem{PhysRevB.82.045102}
Q.~Liu, H.~Yao, and T.~Ma, Phys.~Rev.~B {\bf 82}, 045102 (2010).

\bibitem{PhysRevB.82.075125}
J.~Wen, and A.~R\"uegg, C.-C.~J.~Wang, and G.~A.~Fiete, Phys.~Rev.~B {\bf 82}, 075125 (2010).
\bibitem{doi:10.7566/JPSJ.83.014703}
Y.~Yanase, J.~Phys.~Soc.~Jpn. {\bf 83}, 014703 (2014).

\bibitem{doi:10.7566/JPSJ.83.114704}
T.~Hitomi and Y.~Yanase, J.~Phys.~Soc.~Jpn. {\bf 83}, 114704 (2014).
\bibitem{PhysRevLett.109.237207}
H.~Ishizuka and Y.~Motome, Phys.~Rev.~Lett. {\bf 109}, 237207 (2012).

\bibitem{PhysRevB.88.100402}
H.~Ishizuka and Y.~Motome, Phys.~Rev.~B {\bf 88}, 100402 (2013).

\bibitem{PhysRevB.91.155132}
Y.~Akagi and Y.~Motome, Phys.~Rev.~B {\bf 91}, 155132 (2015).
\end{thebibliography}
\end{document}